\documentclass[twocolumn,english,aps,onecolumn,preprint]{revtex4}
\usepackage[T1]{fontenc}
\usepackage[latin9]{inputenc}
\usepackage{float}
\usepackage{textcomp}
\usepackage{amsthm}
\usepackage{amsmath}
\usepackage{graphicx}
\usepackage{amssymb}
\usepackage{esint}

\makeatletter
\@ifundefined{textcolor}{}
{%
 \definecolor{BLACK}{gray}{0}
 \definecolor{WHITE}{gray}{1}
 \definecolor{RED}{rgb}{1,0,0}
 \definecolor{GREEN}{rgb}{0,1,0}
 \definecolor{BLUE}{rgb}{0,0,1}
 \definecolor{CYAN}{cmyk}{1,0,0,0}
 \definecolor{MAGENTA}{cmyk}{0,1,0,0}
 \definecolor{YELLOW}{cmyk}{0,0,1,0}
 }

\makeatother

\usepackage{babel}

\begin{document}

\title{Hybrid quantum circuit with a superconducting qubit coupled to a
spin ensemble}

\author{Y. Kubo$^{1}$, C. Grezes$^{1}$, A. Dewes$^{1}$, T. Umeda$^{2}$,
J. Isoya$^{2}$, H. Sumiya$^{3}$, N. Morishita$^{4}$, H. Abe$^{4}$,
S. Onoda$^{4}$, T. Ohshima$^{4}$, V. Jacques$^{5}$, A. Dréau$^{5}$,
J.-F. Roch$^{5}$, I. Diniz$^{6}$, A. Auffeves$^{6}$, D. Vion$^{1}$,
D. Esteve$^{1}$, and P. Bertet$^{1}$}

\affiliation{$^{1}$Quantronics group, SPEC (CNRS URA 2464), IRAMIS, DSM, CEA-Saclay,
91191 Gif-sur-Yvette, France }

\affiliation{$^{2}$ Research Center for Knowledge Communities, University of
Tsukuba, Tsukuba 305-8550, Japan}

\affiliation{$^{3}$Sumitomo Electric Industries Ltd., Itami 664-001, Japan}

\affiliation{$^{4}$ Japan Atomic Energy Agency, Takasaki 370-1292, Japan}

\affiliation{$^{5}$ LPQM (CNRS UMR 8537), ENS de Cachan, 94235 Cachan, France}

\affiliation{$^{6}$ Institut Néel, CNRS, BP 166, 38042 Grenoble, France }

\date{\today}

\maketitle
\textbf{Present-day implementations of quantum information processing
rely on two widely different types of quantum bits (qubits). On the
one hand, microscopic systems such as atoms or spins are naturally
well decoupled from their environment and as such can reach extremely
long coherence times \cite{Roos2004,NVLongCoherence}; on the other
hand, more macroscopic objects such as superconducting circuits are
strongly coupled to electromagnetic fields, making them easy to entangle
\cite{diCarlo2010,Neeley2010} although with shorter coherence times
\cite{transmon_exp,Paik2011}. It thus seems appealing to combine
the two types of systems in hybrid structures that could possibly
take the best of both worlds. Here we report the first experimental
realization of a hybrid quantum circuit in which a superconducting
qubit of the transmon type \cite{transmon_th,transmon_exp} is coherently
coupled to a spin ensemble consisting of nitrogen-vacancy (NV) centers
in a diamond crystal \cite{Jelezko2004} via a frequency-tunable superconducting
resonator \cite{TunableResonatorsPalacios} acting as a quantum bus.
Using this circuit, we prepare arbitrary superpositions of the qubit
states that we store into collective excitations of the spin ensemble
and retrieve back later on into the qubit. We demonstrate that this
process preserves quantum coherence by performing quantum state tomography
of the qubit. These results constitute a first proof of concept of
spin-ensemble based quantum memory for superconducting qubits \cite{HybridsImamoglu,HybridsMolmer,HybridMarcos}.
As a landmark of the successful marriage between a superconducting
qubit and electronic spins, we detect with the qubit the hyperfine
structure of the NV center.}

\begin{figure}
\includegraphics[width=160mm]{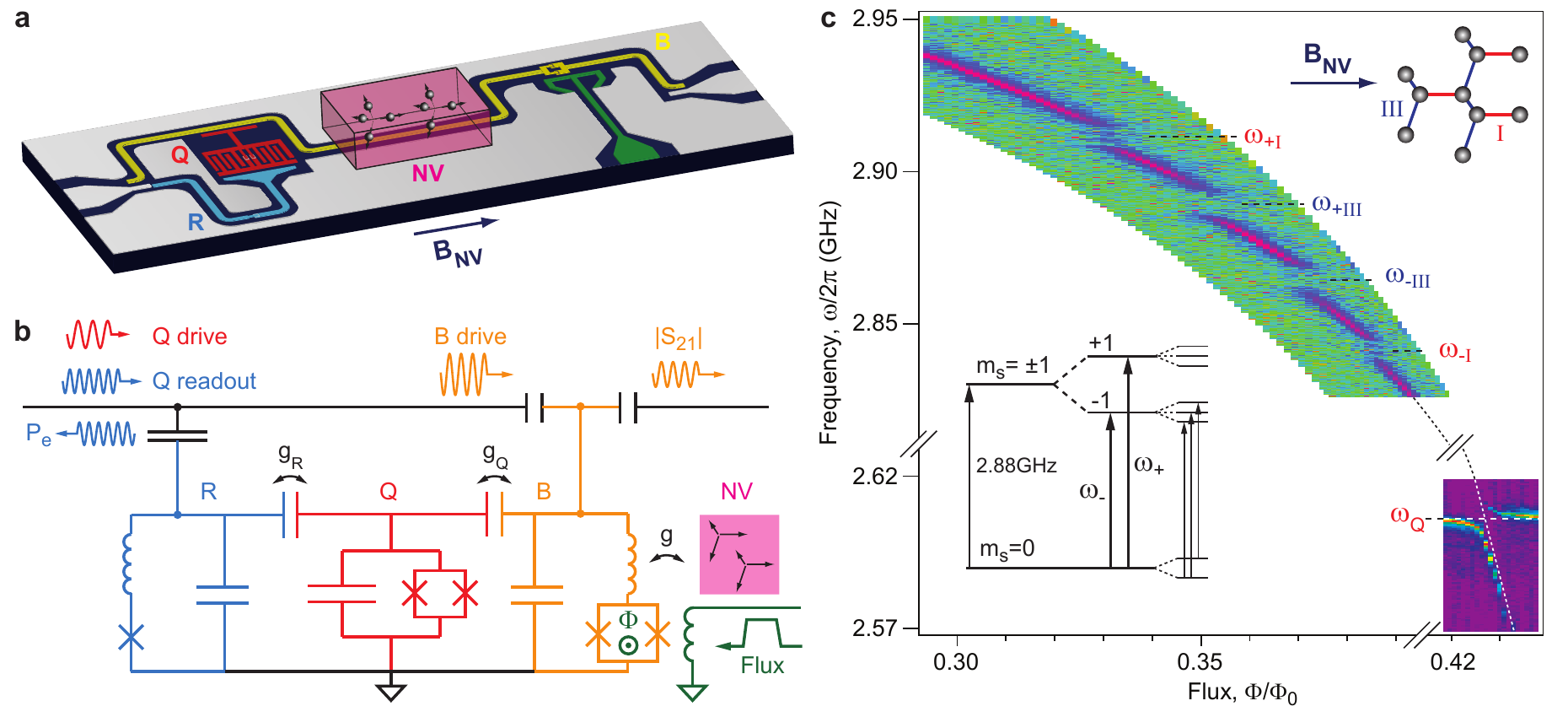}

\caption{Description of the hybrid quantum circuit demonstrated in this work.
\textbf{a},\textbf{b} Three-dimensional sketch of the device and corresponding
electrical scheme. The ensemble $NV$ of electronic spins (magenta)
consists of $10^{12}$ NV centers in a diamond crystal glued on the
chip surface. The transmon qubit $Q$ (in red) is capacitively coupled
to a resonator $R$ (in blue) made nonlinear with a Josephson junction
and used to read-out its state. The bus $B$ (in yellow) is electrostatically
coupled to $Q$ and magnetically coupled to $NV$. $B$ contains a
SQUID that makes its frequency $\omega_{B}(\Phi)$ tunable by changing
the flux $\Phi$ in the SQUID loop applied via a fast on-chip current
line (in green). A magnetic field $B_{NV}$ is applied parallel to
the $[1,1,1]$ crystallographic axis. \textbf{c}, (lower left inset)
Energy level structure of NV centers. Transitions between $m_{S}=0$
and $m_{S}=\pm1$ at frequency $\omega_{\pm}$ are further split in
three resonance lines due to the hyperfine interaction with the $^{14}N$
nuclear spin \cite{Felton2009}. (main panel) Two-dimensional plot
of the transmission $\left|S_{21}\right|(\omega,\Phi)$ through $B$
in dB units, with $\Phi$ expressed in units of the superconducting
flux quantum $\Phi_{0}=h/2e$, for a field $B_{NV}=1.4$~mT applied
to the spins. Color scale goes from $-55$~dB (green) to $-30$~dB
(magenta). Four vacuum Rabi splittings are observed whenever $\omega_{B}$
matches one NV center resonance frequency. The four frequencies correspond
to the $\omega_{\pm I,III}$ transition frequencies of one of two
distinct families of NV centers, being either along the $[1,1,1]$
crystal direction parallel to $B_{NV}$ ($I$, in red), or along one
of the three other possible $\left\langle 1,1,1\right\rangle $ axes
(labelled $III$, in blue), as shown in upper right panel. (main panel,
bottom right) Qubit excited state probability $P_{e}$ as a function
of the frequency of the exciting microwave and $\Phi$. Color scale
goes from $0.1$ (purple) to $0.3$ (yellow). When $\omega_{B}$ matches
the qubit frequency $\omega_{Q}=2.607$~GHz, the qubit spectrum shows
an anticrossing demonstrating its coupling to $B$ with constant $g_{Q}/2\pi=7.2$~MHz.}

\label{fig1} %
\end{figure}

Superconducting qubits have been successfully coupled to electromagnetic
\cite{Wallraff} as well as mechanical \cite{OConnell2010} resonators;
but coupling them to microscopic systems in a controlled way has up
to now remained an elusive perspective - even though qubits sometimes
turn out to be coupled to unknown and uncontrolled microscopic degrees
of freedom with relatively short coherence times \cite{Neeley2008}.
Whereas the coupling constant $g$ of one individual microscopic system
to a superconducting circuit is usually too weak for quantum information
applications, ensembles of $N$ such systems are coupled with a constant
$g\sqrt{N}$ enhanced by collective effects. This makes possible to
reach a regime of strong coupling between one collective variable
of the ensemble and the circuit. This collective variable, which behaves
in the low excitation limit as a harmonic oscillator, has been proposed
\cite{HybridsImamoglu,HybridsMolmer,HybridMarcos} as a quantum memory
for storing the state of superconducting qubits. Experimentally, the
strong coupling between an ensemble of electronic spins and a superconducting
resonator has been demonstrated spectroscopically \cite{Kubo2010,Schuster2010,Majer2011},
and the storage of a microwave field into collective excitations of
a spin ensemble has been observed very recently \cite{BriggsEnsemble,Kubo2011}.
These experiments were however carried out in a classical regime since
the resonator and spin ensemble behaved as two coupled harmonic oscillators
driven by large microwave fields. In the perspective of building a
quantum memory, it is instead necessary to perform experiments at
the level of a single quantum of excitation. For that purpose, we
integrate for the first time on the same chip three different quantum
systems : an ensemble of electronic spins, a superconducting qubit,
and a resonator acting as a quantum bus between the qubit and the
spins. A sketch of the experiment is shown in Fig. \ref{fig1}. 

The spin ensemble $NV$ consists of $\sim10^{12}$ negatively-charged
NV color centers \cite{Jelezko2004} in a diamond crystal. These centers
have an electronic spin $S=1$, with electron spin resonance (ESR)
transition frequencies $\omega_{\pm}$ between energy levels $m_{S}=0$
and $m_{S}=\pm1$ of about $2.88$~GHz in zero magnetic field (see
Fig. \ref{fig1}c). The electronic spin of the NV center is further
coupled by hyperfine (HF) interaction to the spin-one $^{14}N$ nucleus,
which splits $\omega_{\pm}$ into three peaks separated by $2.2$~MHz
\cite{Felton2009} (see Supplementary Material). In our experiment,
the diamond crystal is glued on top of the chip, and the degeneracy
between states $m_{S}=\pm1$ is lifted with a $B_{NV}=1.4$~mT magnetic
field applied parallel to the chip and along the $[1,1,1]$ crystalline
axis. The NV frequencies being sensitive only to the projection of
$B_{NV}$ along the $N-V$ axis, two groups of NVs thus experience
different Zeeman effects: those along $[1,1,1]$ (denoted $I$) and
those along either of the three other $\left\langle 1,1,1\right\rangle $
axes (denoted $III$ as they are $3$ times more numerous). This results
in four different ESR frequencies $\omega_{\pm I,\pm III}$. 

The qubit $Q$ is a Cooper-pair box of the transmon type \cite{transmon_exp,transmon_th}
with transition frequency $\omega_{Q}$ between its ground state $\left|g\right\rangle $
and excited state $\left|e\right\rangle $. It is coupled to a nonlinear
resonator $R$ which is used to read-out its state, as in related
circuit quantum electrodynamics experiments \cite{Mallet2009}. Single-qubit
rotations are realized by applying microwave pulses at $\omega_{Q}$
through $R$. Qubit state readout is performed by measuring the phase
of a microwave pulse reflected on $R$, which depends on the qubit
state; the probability $P_{e}$ to find the qubit in its excited state
is then determined by repeating $\sim10^{4}$ times the same experimental
sequence. 

The quantum bus $B$, a superconducting resonator with quality factor
$\sim10^{4}$, is electrostatically coupled to the qubit and magnetically
coupled to the spin ensemble. In order to bridge the difference in
frequency between $Q$ and $NV$, the bus frequency $\omega_{B}$
can be tuned on a nanosecond time scale \cite{Sandberg} by applying
current pulses through an on-chip line, inducing a magnetic flux $\Phi$
through a SQUID embedded in $B$ \cite{TunableResonatorsPalacios}.
More information on the qubit readout and setup can be found in the
Supplementary Material.

We first characterize our hybrid circuit by spectroscopic measurements.
The NV frequencies and coupling constants are obtained by measuring
the microwave transmission $\left|S_{21}(\omega)\right|$ through
the bus, while scanning its frequency $\omega_{B}(\Phi)$ across the
NV resonance. Vacuum Rabi splittings are observed when $\omega_{B}$
matches the spin resonance frequency at $\omega_{+I}/2\pi=2.91$~GHz,
$\omega_{-I}/2\pi=2.84$~GHz, $\omega_{+III}/2\pi=2.89$~GHz, and
$\omega_{-III}/2\pi=2.865$~GHz (see Fig.~\ref{fig1}). From the
data we extract the coupling constants $g_{\pm I}/2\pi=2.9$~MHz
and $g_{\pm III}/2\pi=3.8$~MHz, the difference between the two values
resulting essentially from the larger number of NV centers in group
$III$. Qubit spectroscopy is performed by scanning the frequency
of a microwave pulse applied through $R$, and by measuring $P_{e}$,
which yields $\omega_{Q}/2\pi=2.607$~GHz. This spectroscopy, measured
while scanning $\omega_{B}$ across $\omega_{Q}$, shows an anticrossing
(see Fig. \ref{fig1}c) that yields the coupling constant $g_{Q}/2\pi=7.2$~MHz
between $Q$ and $B$. 

Throughout the experiments reported in the following, the spins and
qubit frequencies are kept fixed, and only $\omega_{B}$ is varied
in order to transfer coherently quantum information between $Q$ and
$NV$. For this purpose, a key operation is the qubit-bus SWAP gate
that transfers an arbitrary qubit state $\alpha\left|g\right\rangle +\beta\left|e\right\rangle $
into the corresponding photonic state $\alpha\left|0\right\rangle _{B}+\beta\left|1\right\rangle _{B}$
of the bus, leaving the qubit in $\left|g\right\rangle $. This SWAP
gate could be performed by tuning $\omega_{B}$ in resonance with
$\omega_{Q}$ for a duration $\pi/2g_{Q}$ \cite{hofheinz2008}. Here
we prefer instead to adiabatically sweep $\omega_{B}$ across $\omega_{Q}$
as this sequence is more immune to flux noise in the SQUID loop of
$B$ (see Supplementary Material). This adiabatic SWAP ($aSWAP$)
achieves the same quantum operation as the resonant SWAP except for
an irrelevant dynamical phase. The experiments then proceed by combining
single-qubit rotations, $aSWAP$ gates, and flux pulses placing $B$
and $NV$ in and out of resonance for properly chosen interaction
times.

\begin{figure}
\includegraphics[width=89mm]{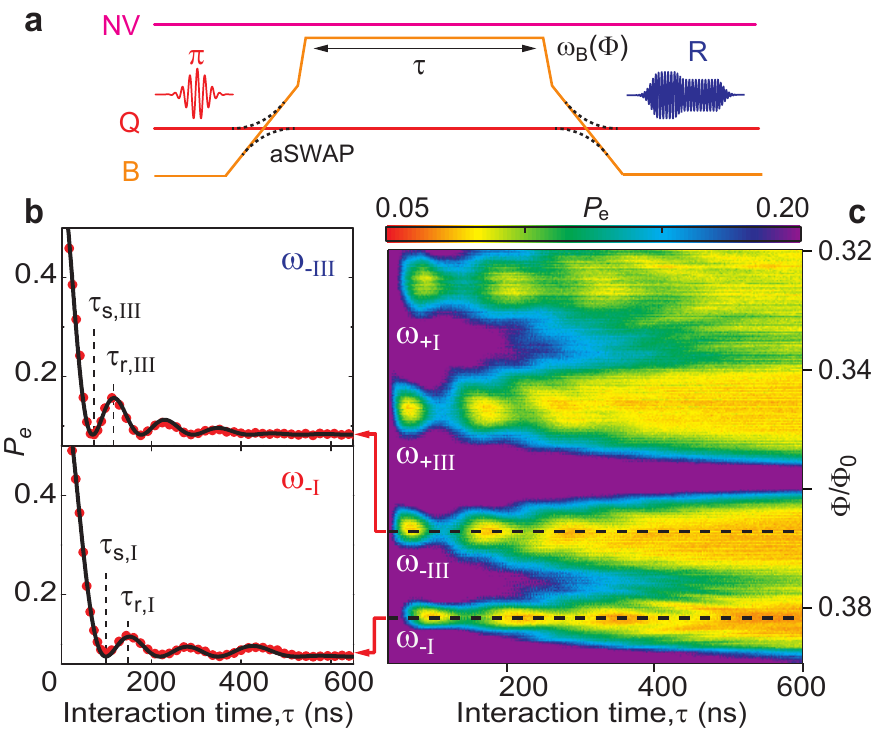}

\caption{Storage and retrieval of a single quantum of excitation from the qubit
to the spin ensemble. \textbf{a}, Experimental sequence showing the
microwave pulses used for exciting the qubit in $\left|e\right\rangle $
(red) and for reading it out (blue), as well as transition frequencies
of the quantum bus (orange), qubit (red), and spins (magenta). \textbf{b},
Experimental (red dots) and theoretical (black line - see text) probability
$P_{e}(\tau)$ for $\omega_{B}$ tuned to $\omega_{-III}$ (top graph)
or $\omega_{-I}$ (bottom graph), showing the storage and retrieval
times $\tau_{s}$ and $\tau_{r}$. \textbf{c}, Two-dimensional plot
of $P_{e}$ versus interaction time $\tau$ and flux pulse height
$\Phi$, showing resonance with the four spin groups. Chevron-like
patterns are observed, showing a faster oscillation with reduced amplitude
when $\omega_{B}$ is detuned from the spin resonance, as expected.
Note that the difference between the $\omega_{-}$ and $\omega_{+}$
patterns in the same NV group is simply caused by the non-linear dependence
of $\omega_{B}$ on $\Phi$ \cite{TunableResonatorsPalacios}.}

\label{fig2} %
\end{figure}

We apply such a sequence with the qubit initially prepared in $\left|e\right\rangle $
(see Fig. \ref{fig2}). A first $aSWAP$ converts $\left|e\right\rangle $
into the bus Fock state $\left|1\right\rangle _{B}$; $B$ is brought
in or near resonance with the spin ensemble for a duration $\tau$;
the resulting $B$ state is then transferred back into the qubit,
which is finally read-out. Figure \ref{fig2}b shows the resulting
curve $P_{e}(\tau)$ when the bus is brought in resonance either with
$\omega_{-III}$ or $\omega_{-I}$. An oscillation in $P_{e}$ is
observed, revealing a storage in the spin ensemble of the single quantum
of excitation initially in the qubit at $\tau_{s,III}=65$~ns or
$\tau_{s,I}=97$~ns, and a retrieval back into the qubit at $\tau_{r,III}=116$~ns
or $\tau_{r,I}=146$~ns. The fidelity of this storage-retrieval process,
defined as $P_{e}(\tau_{r})/P_{e}(0)$, is $0.14$ for group $III$
and $0.07$ for group $I$. These relatively low values are not due
to a short spin dephasing time, but rather to an interference effect
caused by the HF structure of NV centers, as evidenced by the non-exponential
damping observed in $P_{e}(\tau)$. These measurements are accurately
reproduced by a full calculation of the spin-resonator dynamics (see
Suppl. Mat. and \cite{Kurucz2011,Diniz2011,Kubo2011}) taking into
account this HF structure, with the linewidth of each HF peak as the
only adjustable parameter. A linewidth of $1.6$~MHz is in this way
determined for the spins in group $I$, and of $2.4$~MHz for group
$III$, this larger value being likely due to a residual misalignment
of $B_{NV}$ from the $[1,1,1]$ crystalline axis causing each of
the three $<1,1,1>$ $N-V$ orientations non-collinear with the field
to experience slightly different Zeeman shifts. We finally note that
in both curves shown in Fig. \ref{fig2}b $P_{e}(\tau)$ tends towards
$0.08$ at long times, as is also found with the qubit initially in
$\left|g\right\rangle $. This proves that the collective spin variable
coupled to $B$ is, as requested for experiments in the quantum regime,
in its ground state $\left|0\right\rangle _{-I,-III}$ with a large
probability $\sim0.92$ at equilibrium, which corresponds to a temperature
of $\sim50$~mK. Varying both $\omega_{B}$ and $\tau$ with the
same pulse sequence, we observe similar storage-retrieval cycles at
all four spin frequencies (see Fig. \ref{fig2}c).

\begin{figure}
\includegraphics[width=89mm]{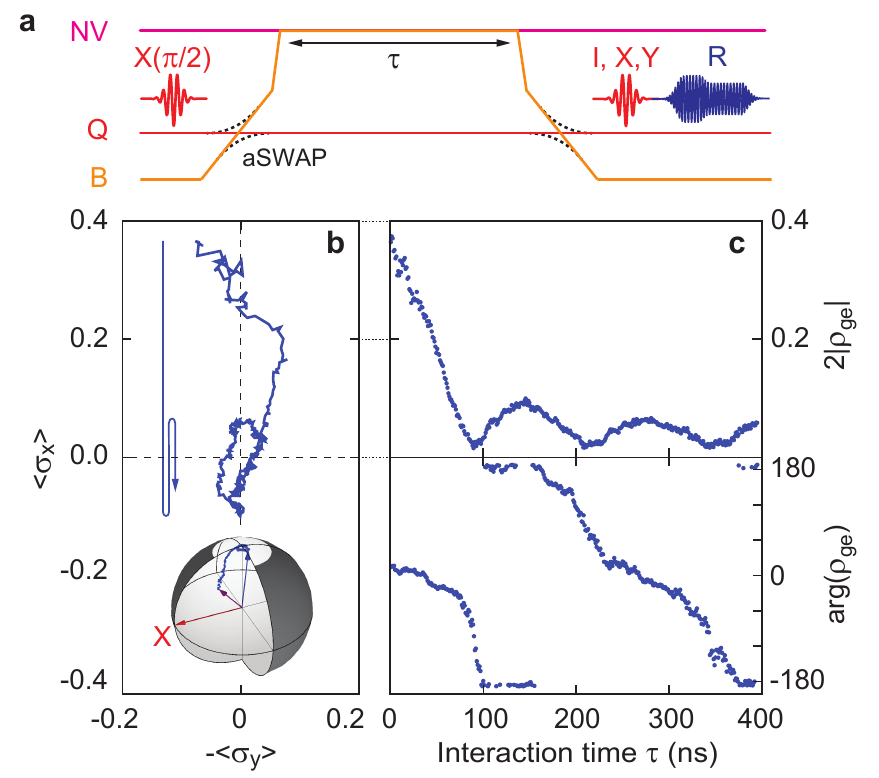}

\caption{Storage and retrieval of a coherent superposition of states from the
qubit to the spin ensemble. \textbf{a}, Experimental pulse sequence:
the qubit is prepared by a $\pi$/2 pulse in state $(|g\rangle+|e\rangle)/\sqrt{2}$
, which is transferred to $B$ by an $aSWAP$. $B$ is then immediately
tuned to $\omega_{-\mathrm{I}}/2\pi=2.84$~GHz for an interaction
time $\tau$. The quantum state of $B$ is then transferred back to
the qubit by an $aSWAP$. Quantum state tomography is finally performed
to determine the qubit state by applying either $I$, $X$, or $Y$
operation to the qubit. \textbf{b, }Trajectory of the qubit Bloch
vector on the Bloch sphere (bottom inset), and its projection on the
equatorial plane. \textbf{c,} Modulus and phase of the off-diagonal
element $\rho_{ge}$ of the qubit density matrix as a function of
interaction time $\tau$.}

\label{fig3} %
\end{figure}

In addition to storing a single excitation from the qubit, one has
to test if a coherent superposition of states can be transferred to
the spin ensemble and retrieved. For that, we now perform the $aSWAP$
and bring $\omega_{B}$ in resonance with $\omega_{-I}$ after having
initialized the qubit in $(\left|g\right\rangle +\left|e\right\rangle )/\sqrt{2}$
instead of $\left|e\right\rangle $, and we reconstruct the Bloch
vector of the qubit by quantum state tomography at the end of the
sequence. More precisely, we measure $\left\langle \sigma_{X}\right\rangle $,
$\left\langle \sigma_{Y}\right\rangle $ and $\left\langle \sigma_{Z}\right\rangle $
by using $\pi/2$ rotations around $Y$, $X$, or no rotation at all
($I$) prior to qubit readout. After substracting a trivial rotation
around $Z$ occurring at frequency $\left(\omega_{-I}-\omega_{Q}\right)$,
we reconstruct the trajectory of this Bloch vector as a function of
the interaction time $\tau$. It is plotted in Fig. \ref{fig3}, together
with the off-diagonal element $\rho_{ge}$ of the final qubit density
matrix, which quantifies its coherence. We find that no coherence
is left in the qubit at the end of the sequence for $\tau=\tau_{s,I}$,
as expected for a full storage of the initial state into the ensemble.
Then, coherence is retrieved at $\tau=\tau_{r,I}$, although with
an amplitude $\sim5$ times smaller than its value at $\tau=0$ (i.e.
without interaction with the spins). Note the $\pi$ phase shift occurring
after each storage-retrieval cycle, characteristic of $2\pi$ rotations
in the two-level space $\left\{ \left|1_{B},0_{-I}\right\rangle ,\left|0_{B},1_{-I}\right\rangle \right\} $.
The combination of the results of Figs. \ref{fig2} and \ref{fig3}
demonstrates that arbitrary superpositions of the two qubit states
can be stored and retrieved in a spin ensemble - although with limited
fidelity - and thus represents a first proof-of-concept of a spin-based
quantum memory for superconducting qubits. 

\begin{figure}[t]
\includegraphics[width=85mm]{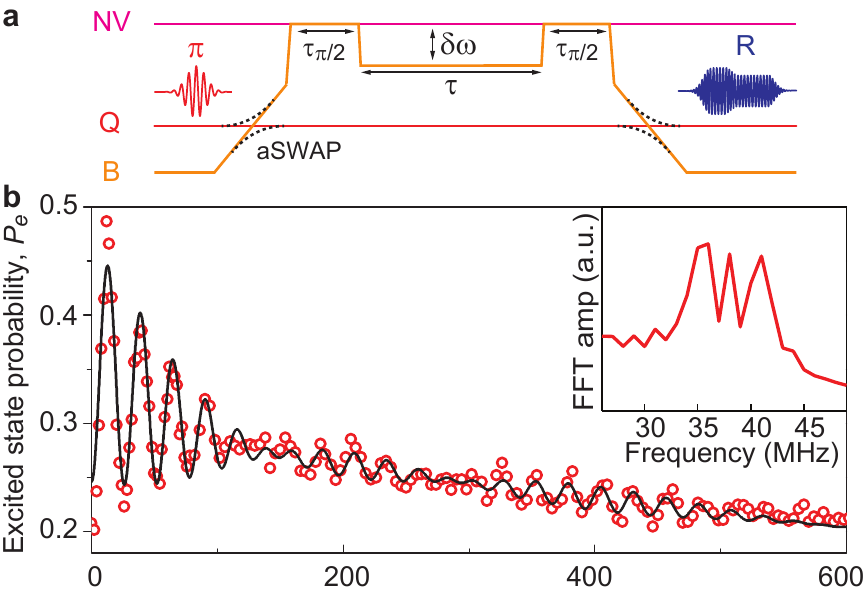}

\caption{Ramsey-like experiment on the spin ensemble at the single-photon level.
\textbf{a}, Experimental pulse sequence: the qubit is prepared in
its excited state $|e\rangle$ by a $\pi$ pulse; the state $\left|e,0_{B}\right\rangle $
is then adiabatically transferred to $\left|g,1_{B}\right\rangle $
by an $aSWAP$. A fast flux pulse subsequently brings $\omega_{B}$
onto $\omega_{-\mathrm{I}}$, and then lets $B$ and the spins from
group $-\mathrm{I}$ interact for half a swap time $\tau_{\mathrm{s,I}}/2$,
generating an entangled state of the two systems. $B$ is then detuned
from the spins by $\delta\omega/2\pi=38$ MHz during a time $\tau$,
and a second half-swap is performed. The quantum state of $B$ is
then transferred back to the qubit, which is finally read-out. \textbf{b},
Measured (red circles) and calculated (black line - see text) probability
$P_{e}(\tau)$, as well as its Fourier transform (inset) revealing
the NV centers HF structure.}

\label{fig4} %
\end{figure}

To evaluate the time during which quantum coherence can be stored
in the ensemble, we perform a Ramsey-like experiment on the spin ensemble
at the single-photon level (see Fig. \ref{fig4}): we initially prepare
the qubit in $\left|e\right\rangle $, transfer its state to $B$,
then tune $\omega_{B}$ to $\omega_{-I}$ for a duration $\tau_{\pi/2}=\tau_{s,-I}/2$,
after which $\omega_{B}$ is suddenly detuned by $\delta\omega/2\pi=38$~MHz
for a time $\tau$. At this point, the joint bus-spin ensemble state
is an entangled state $\left(\left|1_{B},0_{-I}\right\rangle +e^{i\varphi}\left|0_{B},1_{-I}\right\rangle \right)/\sqrt{2}$
with a phase $\varphi=\delta\omega\tau$. $B$ is then put back in
resonance with the spins for a second interaction of duration $\tau_{\pi/2}$
that converts the phase $\varphi$ into population of $\left|1_{B},0_{-I}\right\rangle $.
This population is finally transferred to the qubit, and read-out.
Oscillations at frequency $\delta\omega$ are observed in $P_{e}(\tau)$
as seen in Fig. \ref{fig4}, confirming that the resonator and the
spins are entangled after the first $\pi/2$ pulse. These oscillations
are modulated by a beating pattern, with an overall damping of the
oscillations envelope in $\sim200$~ns. Quite remarkably, this beating
observed in the qubit excited state probability is directly caused
by the HF structure of NV centers, as proved by the Fourier transform
of $P_{e}(\tau)$ which shows the three HF lines. The full calculation
of the system dynamics quantitatively captures both the beatings and
the oscillations damping, which is thus completely explained by the
$1.6$~MHz inhomogeneous linewidth of each HF line taken into account
in the theory. 

The previous results suggest that the storage of quantum information
in the NV centers ensemble is at present limited both by its HF structure
and by the inhomogeneous broadening of its resonance. This broadening
is attributed to dipolar interactions between the NV centers and residual
paramagnetic impurities (likely neutral nitrogen atoms) in the diamond
crystal. Purer crystals could thus greatly improve the present performance
of our device. Note that the hyperfine coupling to the nuclear spin
of $^{14}N$ could be turned into a useful resource if quantum information
was transferred from the electron spin to the nuclear spin degree
of freedom, which has much narrower linewidth. Finally refocusing
techniques borrowed from quantum memories in the optical domain \cite{Lvovsky2009}
should also lead to increase in the storage time by two orders of
magnitude. 

In conclusion our experiments bring the first proof of concept of
a spin-based quantum memory for superconducting qubits. In a longer-term
perspective, they open the way to the implementation of genuine quantum
lab-on-chips, where superconducting qubits would coherently interact
with electron and nuclear spins as well as optical photons. 

\textbf{Note:} During redaction of this manuscript we became aware
of related work demonstrating the coherent dynamics of a flux qubit
coupled to an ensemble of NV centers in diamond \cite{Zhu2011}.

\textbf{Acknowledgements} We acknowledge useful discussions with K.
Moelmer, F. Jelezko, J. Wrachtrup, D. Twitchen, and within the Quantronics
group, and technical support from P. Sénat, P.-F. Orfila, T. David,
J.-C. Tack, P. Pari, P. Forget, M. de Combarieu. We acknowledge support
from European projects Midas and Solid, ANR project Masquelspec, C'Nano,
Capes, and Fondation Nanosciences de Grenoble. 

\textbf{Methods Summary}

\textbf{Superconducting circuit parameters} The transmon parameters
are measured by standard spectroscopy, yielding a Josephson energy
$E_{J}/h=5.2$~GHz and a Coulomb energy for a Cooper-pair $E_{C}/h=0.66$~GHz.
Its relaxation time $T_{1}=1.75$~$\mu\mathrm{s}$ and coherence
time $T_{2}=2.2$~$\mu\mathrm{s}$ at the bias point used in this
work were measured by standard pulse sequences. 

The bus resonator $B$ could be tuned from a maximum frequency $\omega_{B}(0)/2\pi=3.004$~GHz
down to $2.5$~GHz. Its quality factor is $2\cdot10^{4}$ at $\Phi=0$
in presence of the diamond sample (which thus does not introduce dielectric
losses contrary to what was reported in \cite{Kubo2010,Majer2011}).
This quality factor degrades progressively as $\omega_{B}$ is tuned
towards lower frequencies. However the bus resonator energy relaxation
time $T_{cav}=1.5$~$\mu\mathrm{s}$ was measured using the qubit
as explained in \cite{Wang2008}, and was found not to depend on the
flux bias. This indicates that the quality factor degradation is due
to low-frequency noise, likely flux noise in the SQUID loop.

The readout resonator $R$ has a frequency $\omega_{R}/2\pi=3.468$~GHz
and a quality factor $Q=500$. Its nonlinearity is brought by a Josephson
junction of critical current $650$~nA yielding a Kerr constant $K/\omega_{R}=-4.5\cdot10^{-6}$
\cite{Ong2011}. Readout pulses have a frequency $3.456$~GHz. The
coupling between $Q$ and $R$ is estimated to be $g_{R}/2\pi=30$~MHz. 

\textbf{Theory} Each spin is modelled by an effective harmonic oscillator
$b_{j}$ of frequency $\omega_{j}$, following the Holstein-Primakoff
approximation valid in the low-excitation limit. The qubit, bus resonator,
and spin ensemble are described by Hamiltonians $H_{Q}/\hbar=-(\omega_{Q}/2)\sigma_{Z}$,
$H_{B}/\hbar=\omega_{B}(\Phi)a^{\dagger}a$ and $H_{NV}/\hbar=\sum\omega_{j}b_{j}^{\dagger}b_{j}$,
$\sigma_{Z}$ being the Pauli matrix, $a$ being the bus resonator
annihilation operator. Coupling of the qubit to the bus resonator
is described by a Jaynes-Cummings Hamiltonian $H_{Q-B}/\hbar=g_{Q}(\sigma^{+}a+h.c)$
where $\sigma^{+}$ is the qubit raising operator. Coupling between
the bus resonator and the spin ensemble is described by a Tavis-Cummings
Hamiltonian $H_{B-NV}/\hbar=\sum_{j=1}^{N}g_{j}(b_{j}^{\dagger}a+h.c.)$.
This Hamiltonian can be rewritten as $H_{B-NV}/\hbar=\sum_{K=-III,-I,+I,+III}g_{K}\left(b_{K}^{\dagger}a+h.c.\right)$
where $g_{K}$ represents the collective coupling between B and each
of the four spin groups ($\pm I,$$\pm III$) and $b_{K}=(1/g_{K})\sum_{j=1}^{N}g_{\mathrm{j}}b_{\mathrm{j}}$
is the collective excitation annihilation operator. Excited states
of the spin ensemble are defined for each family by applying collective
operators $b_{K}$ and $b_{K}^{\dagger}$ to the ground state $\left|0_{K}\right\rangle $
(for instance $\left|1_{-I}\right\rangle =b_{-I}^{\dagger}\left|0_{-I}\right\rangle $).
The spin-resonator dynamics can be calculated from this model, as
explained in \cite{Diniz2011,Kurucz2011} and in the Supplementary
Material.

\textbf{Adiabatic pulse parameters }Our adiabatic SWAP operation proceeds
as follows: $\omega_{B}$ starts at $2.52$~GHz, is first ramped
up to $2.589$~GHz in $60$~ns, then to $2.643$~GHz in $350$~ns,
then to $2.687$~GHz in $40$~ns. See the Supplementary Information
for more details on the pulse optimization.

\section*{Supplementary information for {}``Hybrid quantum circuit with a
superconducting qubit coupled to a spin ensemble''}

\subsection{Diamond sample preparation and characterization}

The sample we use is a polished $(110)$ plate of dimensions $2.2\times1\times0.5\,\mathrm{mm}^{3}$
taken from a type-Ib HPHT crystal which contained $40$~ppm of neutral
substitutional nitrogen (the P1 centers) as measured by IR absorption.
It has been irradiated by $2$~MeV electrons at $700\text{\textdegree}\,$C
with a total dose of $6.4\times10^{18}\,\mathrm{e}/\mathrm{cm}^{2}$
and annealed at $1000\text{\textdegree}\,$C for $2$ hours in vacuum.
The high temperature irradiation was employed to minimize the residual
unwanted defects.

\begin{figure}[H]
\centering{}\includegraphics[scale=0.5]{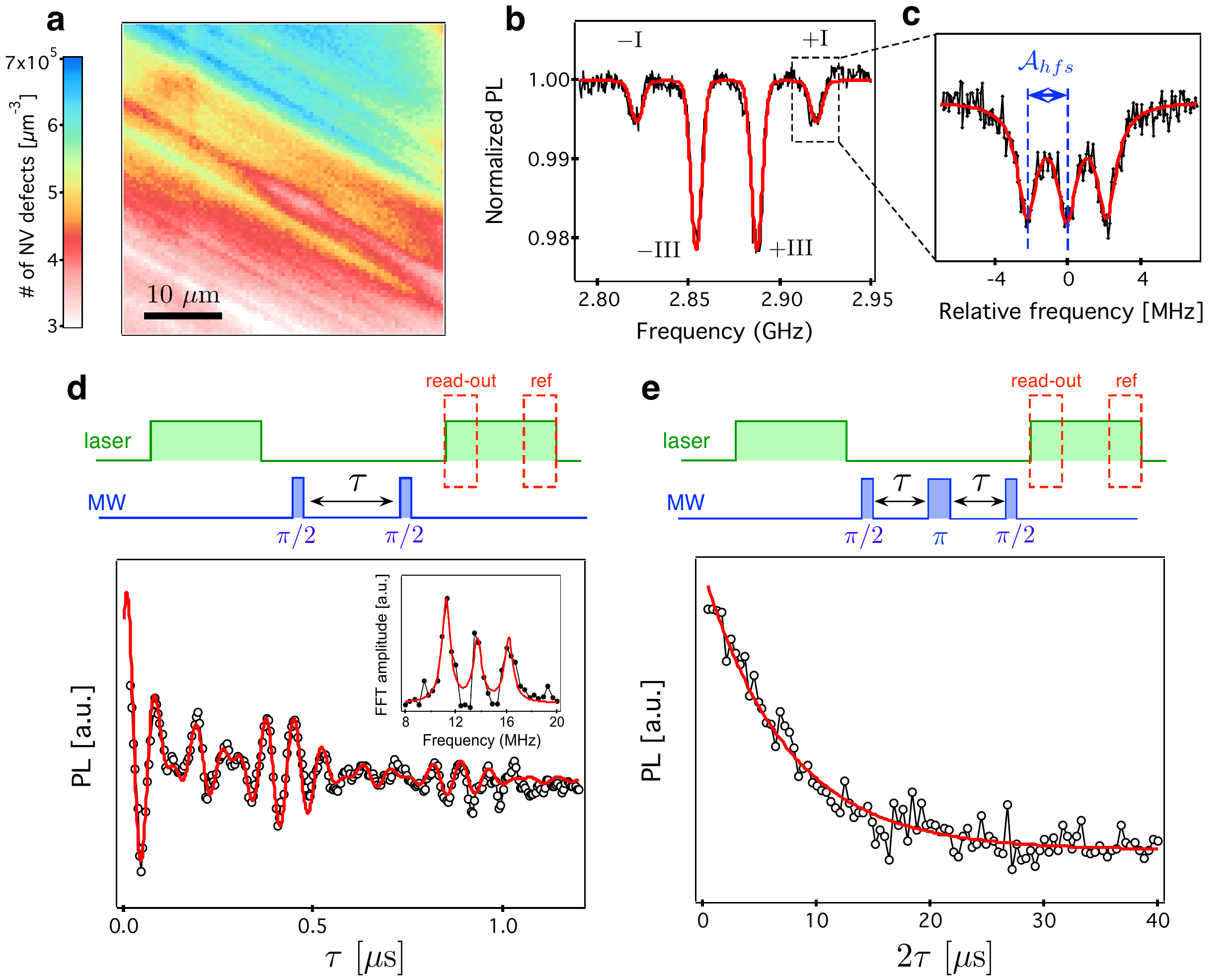} \caption{\textbf{Characterization of the diamond sample at room temperature.}
\textbf{a},Two-dimensional map of the NV center density for a depth
of $5\ \mu$m inside the diamond sample. \textbf{b}, Optically detected
ESR spectrum with a static magnetic field $B\approx1.5$~mT applied
along the {[}111{]} axis of the diamond crystal. Four ESR lines are
observed, corresponding to $m_{s}=0\rightarrow m_{s}=\pm1$ spin transitions
for the two subsets of NV centers crystalline orientations ($\pm{\rm III}$
and $\pm{\rm I}$). \textbf{c}, Hyperfine structure. Data fitting
with Lorentzian functions leads to $\mathcal{A}_{hf}=2.17\pm0.02$
MHz as expected from hyperfine interaction with $^{14}$N nuclear
spins. \textbf{d}, Ramsey fringes recorded for the subset of {[}$111${]}-oriented
NV centers with a microwave detuning $\delta=13$ MHz from the ESR
line at $\omega_{+I}/2\pi=2.915$~GHz. The laser pulses used in the
Ramsey sequence have a duration of $30\ \mu$s and the laser power
is set at $40$~mW. For data analysis, the NV center PL recorded
during the first $10\ \mu$s of the laser pulses is used for spin-state
read-out while the PL recorded \textbf{during} the last $10\ \mu$s
is used as reference. The red solid line is data fitting with the
function $\exp[-\tau/T_{2}^{*}]\times\sum_{i=-1}^{1}\cos\left[2\pi(\delta+i\mathcal{A}_{hf})\tau\right]$.
The inset shows the Fourier-transform of the free induction decay.
Solid lines are data fitting with Lorentzian functions. \textbf{e},
Measurement of the coherence time $T_{2}$ for the subset of {[}$111${]}-oriented
NV centers using a $\pi/2-\tau-\pi-\tau-\pi/2$ spin echo sequence.
Data fitting with an exponential decay leads to $T_{2}=7.3\pm0.4\ \mu$s.\label{FigS1} }
\end{figure}

\indent The resulting concentration of negatively-charged NV centers
was measured by comparing the sample photoluminescence (PL) to the
one obtained from a single NV center. For that purpose, a continuous
laser source operating at $532$~nm wavelength was tightly focused
on the sample through a high numerical aperture oil-immersion microscope
objective. The NV center PL was collected by the same objective, spectrally
filtered from the remaining pump light and directed to a silicon avalanche
photodiode. After calibration of the PL response associated with a
single NV center, a PL raster scan of the sample directly indicates
the NV center content since the excitation volume is known. As shown
in Fig.~\ref{FigS1}(a), the NV center concentration is found rather
inhomogeneous over the sample, with an average density of $4.4\times10^{17}$~cm$^{-3}$.
Such a value is in good agreement with the spin ensemble-resonator
coupling constants reported in the main text of the manuscript.

\indent Electron spin resonance (ESR) measurements were performed
at room temperature by applying a microwave field through a copper
microwire directly spanned on the diamond surface. In addition, a
static magnetic field $B\approx1.5$~mT was applied along the {[}111{]}
axis of the diamond crystal. As explained in the main text, such a
magnetic field orientation allows both to lift the degeneracy of $m_{s}=\pm1$
spin sublevels and to divide the NV center ensemble into two sub-groups
of crystallographic orientations which experience different Zeeman
splittings. Optically detected ESR spectra were recorded by sweeping
the frequency of the microwave field while monitoring the PL intensity.
As shown in Fig.~\ref{FigS1}(b), when the microwave frequency is
resonant with a transition linking $m_{s}=0$ and $m_{s}=\pm1$ spin
sublevels, ESR is evidenced as a dip of the PL signal owing to spin-dependent
PL response of the NV center~\cite{Gruber1997,Manson2006}. In the
following, we focus the study on the ESR line at $\omega_{+I}/2\pi=2.915$~GHz,
which corresponds to the $m_{s}=0\rightarrow m_{s}=+1$ spin transition
for the subset of {[}$111${]}-oriented NV centers. The nitrogen atom
of NV centers in our sample being a $^{14}$N isotope ($99.6\%$ abundance),
corresponding to a nuclear spin $I=1$, each electron spin state is
further split into three sublevels by hyperfine interaction with a
splitting $\mathcal{A}_{hf}=-2.16$ MHz between ESR frequencies associated
with different nuclear spin projections~\cite{Felton2009-1}. This
hyperfine structure can be easily observed in our sample by decreasing
the microwave power in order to reduce power broadening of the ESR
linewidth~\cite{Dreau2011} (Fig.~\ref{FigS1}(c)). \\
 \indent To probe coherence properties of this subset of NV centers,
Ramsey fringes were first recorded by using the usual sequence consisting
of two microwave $\pi/2$-pulses separated by a variable free evolution
duration $\tau$ (Fig.~\ref{FigS1}(d))~\cite{Childdress2006,Bala2009}.
The free induction decay signal exhibits beating frequencies which
correspond to the hyperfine components of the NV center. Data fitting
of the free induction decay signal leads to a dephasing time $T_{2}^{*}=390\pm30$~ns
of the NV center electron spins and its Fourier transform spectrum
reveals the $^{14}$N hyperfine structure with a linewidth (FWHM)
$\Gamma=810\pm90$~kHz for each peak (see inset of Fig.~\ref{FigS1}(d)).\\
 \indent The dephasing time can be greatly enhanced by decoupling
the electron spin from its local environment with a spin echo sequence
(Fig.~\ref{FigS1}(e)). Using this technique, the dephasing time
of the NV center ensemble reaches $T_{2}=7.3\pm0.4\ \mu$s at room
temperature. At high spin densities, this quantity is limited by the
interaction with a bath of paramagnetic impurities including NV centers
themselves and P1 centers~\cite{CoherenceTime,Taylor2008}.

%\addcontentsline{toc}{chapter}{Bibliographie}

\subsection{Superconducting circuit fabrication and measurement setup}

\begin{figure*}
\includegraphics[scale=0.8]{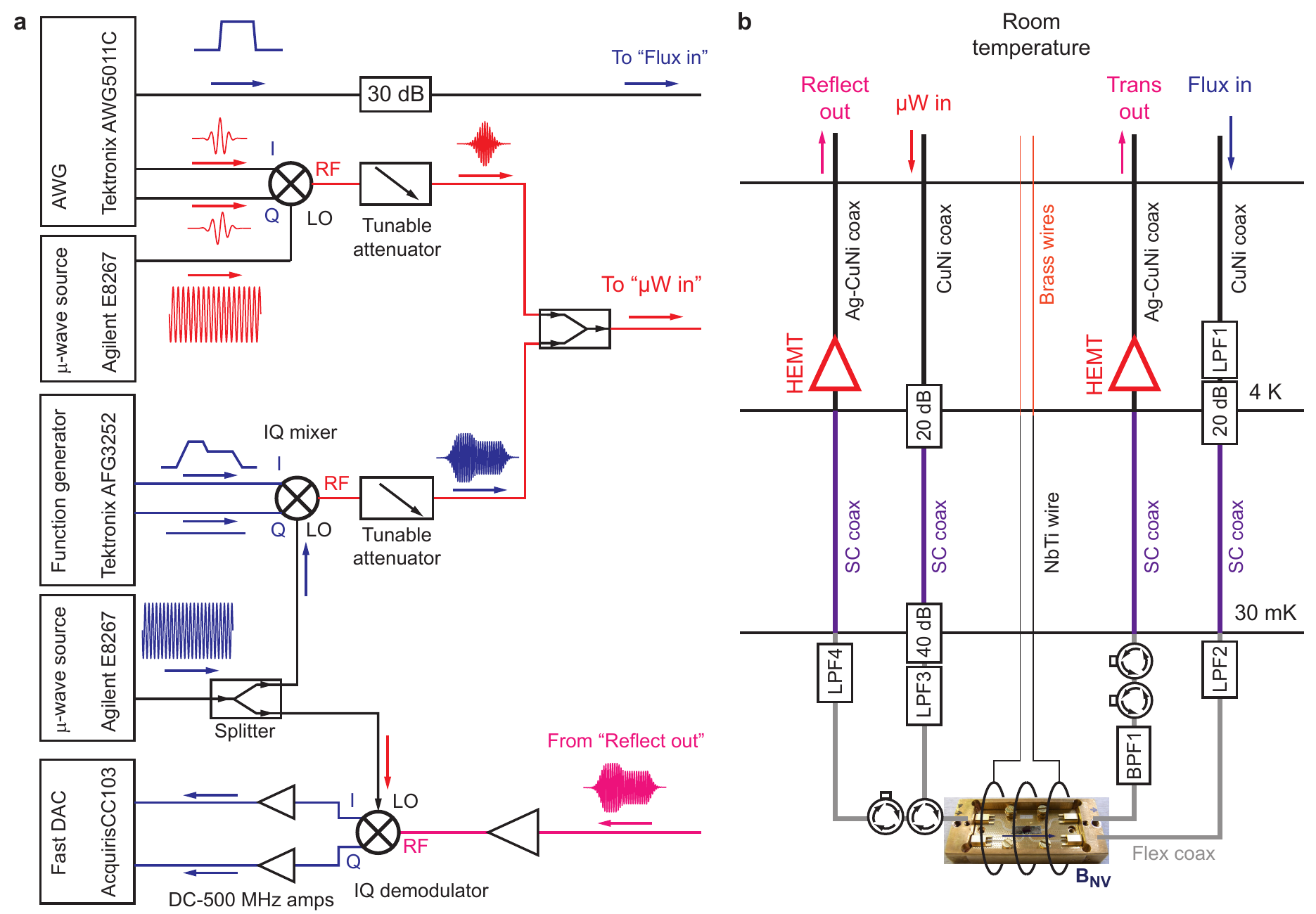}

\caption{Measurement setup and wiring. \textbf{a}, Full configuration of the
measurement apparatus at room temperature. \textbf{b}, Scheme of the
wiring inside the dilution refrigerator. LPF1, LPF2, LPF3, and LPF4
are low-pass filters with cutoff frequencies 1.35 GHz, 450 MHz, 5.4
GHz, and 5.4 GHz, respectively. BPF1 is a band-pass filter with a
bandwidth of 2.5 - 4 GHz. CuNi coax is a coaxial cable made of CuNi,
and Ag-CuNi coax is a silver-plated CuNi coaxial cable. SC coax is
a superconducting NbTi coaxial cable. Flex coax is a low-loss flexible
coaxial cable. Rectangles represent ports terminated by 50 $\Omega$.
The cryogenic microwave amplifier is a CITCRYO 1-12 from Caltech,
with gain $\sim38$~dB and noise temperature $\sim5$~K at $3$~GHz.
A DC magnetic field $B{}_{NV}$ is applied parallel to the chip by
passing a DC current through an outer superconducting coil. The sample
box and the coil are surrounded by two magnetic shieldings consisting
of a lead cylinder around which permalloy tape is wrapped. The sample
box, coil, and the shieldings are thermally anchored at the mixing
chamber.\label{figS2:Setup}}
\end{figure*}

The superconducting circuit is fabricated on a silicon chip oxidized
over $50$~nm. A $150$~nm thick niobium layer is first deposited
by magnetron sputtering and then dry-etched in a $SF_{6}$ plasma
to pattern the readout resonator $R$, the bus resonator $B$, the
current lines for frequency tuning, and the input waveguides. Finally,
the transmon qubit $Q$, the coupling capacitance between $Q$ and
$B$, the Josephson junction of $R$, the SQUID in $B$, are fabricated
by double-angle evaporation of aluminum through a shadow mask patterned
using e-beam lithography. The first layer of aluminum is oxidized
in a $Ar-O_{2}$ mixture to form the oxide barrier of the junctions.
The chip is glued with wax on a printed circuit board (PCB) and wire
bonded to it. The PCB is then screwed in a copper box anchored to
the cold plate of a cryogen-free dilution refrigerator. A complete
scheme of the measurement setup and fridge wiring is shown in Fig.
\ref{figS2:Setup}.

\subsection{Qubit readout}

\begin{figure}[th]
\includegraphics[scale=0.6]{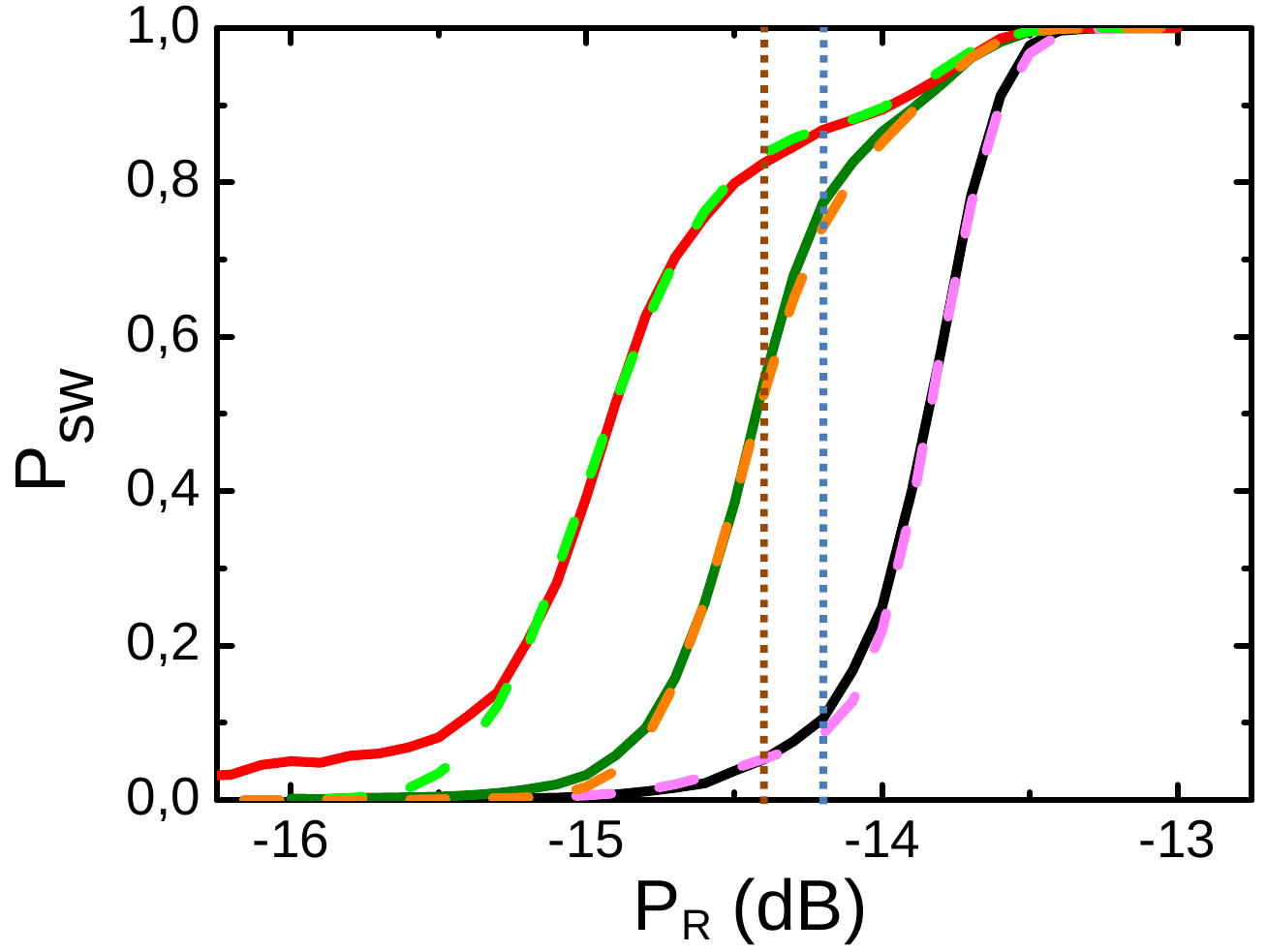}

\caption{Readout resonator switching probability $P_{sw}$ as a function of
readout pulse power $P_{R}$ (S-curves). Black solid line : S-curve
with the qubit in thermal equilibrium. Olive solid line: S-curve with
the qubit prepared in $\left|e\right\rangle $ by a $\pi$ pulse just
before the readout. Red solid line: S-curve with the qubit prepared
in $\left|e\right\rangle $ by a $\pi$ pulse, measured with a composite
readout pulse including a $\pi$ pulse on the $\left|e\right\rangle -\left|f\right\rangle $
transition followed by the usual readout pulse. Dashed lines represent
fits of these S-curves using a sum of three $\mathrm{Erf}$ functions
(corresponding to the three transmon states $\left|g\right\rangle ,\,\left|e\right\rangle ,\,\left|f\right\rangle $)
with different weights. In this way the thermal population of the
qubit $P_{e,eq}=0.08$ is evaluated as explained in the text. Note
that this thermal excitation is responsible for the {}``foot'' of
the black curve. Dotted blue vertical line indicates the readout power
used for simple readout pulses, dotted brown vertical line indicates
the readout power used for the composite readout pulse method.}
\label{FigS3:Scurves}%
\end{figure}

The qubit readout method we use is explained in detail in \cite{Mallet2009-1}.
It relies on the nonlinearity of the readout resonator $R$ operated
in the so-called JBA mode where it behaves as a sample-and-hold detector.
More precisely, we apply a readout pulse of frequency $\omega/2\pi=3.456$~GHz
slightly lower than the resonance frequency $\omega_{R}/2\pi=3.468$~GHz,
and of power $P_{R}$ chosen so that the resonator is driven close
to its bistability, in a regime where the field inside the resonator
can switch from a low-amplitude state $L$ to a high-amplitude state
$H$. This switching is easily detected by measuring the phase of
the reflected readout pulse. Repeating the same sequence then yields
the resonator switching probability $P_{sw}$ for a given readout
pulse power. This allows to reconstruct so-called S-curves $P_{sw}(P_{R})$
which change from $0$ to $1$ in a narrow power range, close to bistability
(see Fig. \ref{FigS3:Scurves}). Due to the qubit-resonator dispersive
coupling, the readout resonator $R$ frequency is shifted by a qubit-state-dependent
quantity $\pm\chi$ so that for well-chosen pulse frequency and power
the qubit state is mapped onto the resonator dynamical state at the
end of the readout pulse (see Fig. \ref{FigS3:Scurves}). The measured
switching probability $P_{sw}$ is therefore directly linked to the
qubit excited state probability $P_{e}$ which is the quantity of
interest in our experiment. 

That $P_{e}$ is not directly given by $P_{sw}$ is due to readout
errors caused either by a too small $\chi$ or by qubit relaxation
between the end of the experimental sequence and the time at which
readout effectively takes place. These errors can be modelled with
two parameters : the probability $e_{0}$ that the resonator switches
despite the qubit being in $\left|g\right\rangle $ at the end of
the experimental sequence, and the probability $e_{1}$ that the resonator
doesn't switch while the qubit is in $\left|e\right\rangle $. In
order to determine $e_{0}$ and $e_{1}$, we measure the switching
probability $P_{sw0}$ for a qubit at thermal equilibrium, and after
a $\pi$ pulse $P_{sw\pi}$ that we assume ideal in the sense that
it swaps states $\left|g\right\rangle $ and $\left|e\right\rangle $
with $100\%$ efficiency. An additional complication arises from the
fact that the qubit has a small but finite probability $P_{e,eq}$
to be found in $\left|e\right\rangle $ even at thermal equilibrium,
due to the rather low qubit frequency chosen in the experiment to
match the NV centers. We therefore first estimate $P_{e,eq}$ by fitting
the shape of S-curves at equilibrium and after a $\pi$ pulse to a
simple model, yielding $P_{e,eq}=0.08$ in our experiment (see Fig.
\ref{FigS3:Scurves}). This corresponds to an effective electromagnetic
temperature of $50$~mK, slightly higher than the cryostat base temperature
$30$~mK possibly due to imperfect filtering of the flux lines. We
then find $e_{0}$ and $e_{1}$ by solving the system of two equations
$P_{sw0}=e_{0}(1-P_{e,eq})+(1-e_{1})P_{e,eq}$ and $P_{sw\pi}=e_{0}P_{e,eq}+(1-e_{1})(1-P_{e,eq})$.
This allows to determine $P_{e}$ from the directly measured $P_{sw}$
since $P_{sw}=e_{0}(1-P_{e})+(1-e_{1})P_{e}$.

An additional complication arises from the fact that the fidelity
of the readout can be enhanced (i.e. $e_{0}$ and $e_{1}$ lowered)
by using the second excited state $\left|f\right\rangle $ of the
transmon: for that, one applies a $\pi$ pulse on the $\left|e\right\rangle -\left|f\right\rangle $
transition just prior to readout, resulting in a so-called composite
readout pulse. As explained in \cite{Mallet2009-1} this reduces readout
errors caused by relaxation during the readout pule. Due to technical
complications, we use the composite readout pulse method only in experiments
reported in figures 2 and 4a of the main article. The other experiments
were performed with simple readout pulses. As a result two different
sets of errors $e_{0}$ and $e_{1}$ were determined for each of the
two types of readout pulses. For composite readout pulses, we find
$e_{0}=0$ and $e_{0}=0.1$, indicating a very high fidelity readout
consistent with \cite{Mallet2009-1}. Without the composite readout
pulse we find $e_{0}=0$ and $e_{0}=0.33$. From these values we convert
the measured $P_{sw}$ into $P_{e}$ in all our experiments.

\subsection{Qubit state manipulation}

\begin{figure}[t]
\includegraphics{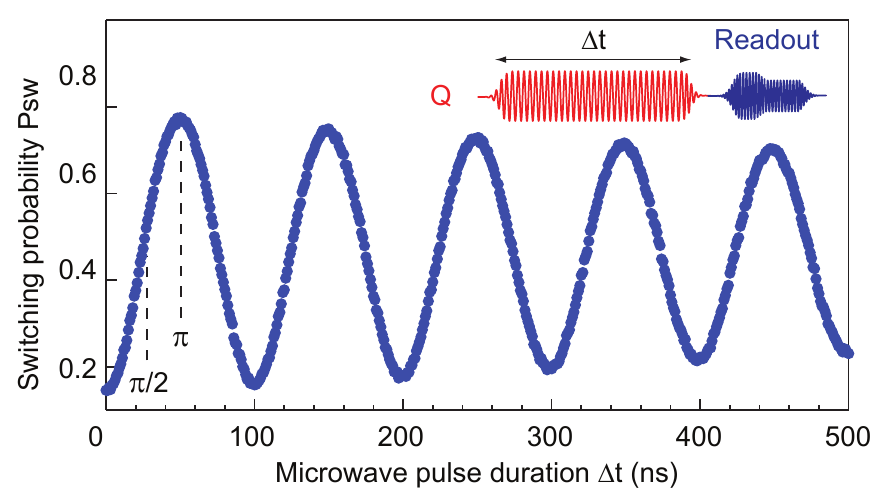}

\caption{Rabi oscillations. The switching probability $P_{SW}$ of readout
resonator $R$ is plotted as a function of microwave pulse duration
$\Delta t$. Inset: pulse sequence used in this measurement. Each
data point of $P_{SW}$ was constructed by repeating this sequence
$10^{4}$ times.}
\label{figS4:Rabi-oscillations}%
\end{figure}

Single-qubit operations are carried out by applying Gaussian shaped
microwave pulses\cite{Lucero} at $\omega_{Q}$. These pulses are
generated as explained above by mixing a CW source at $\omega_{Q}-\omega_{S}$
with a Gaussian shaped pulse modulated at $\omega_{S}$ using an IQ
mixer. Before the measurement such as shown in the main text, the
system was calibrated to compensate the mixer imperfections (amplitude
and phase imbalance, offsets). By changing the sideband frequency
$\omega_{S}$ it is also possible to apply pulses on the $\left|e\right\rangle -\left|f\right\rangle $
transition as requested sometimes for readout. Resulting Rabi oscillations
are shown in \ref{figS4:Rabi-oscillations}. Here the $\pi$ pulse
and $\pi/2$ pulse are defined to be $50$~ns and $25$~ns respectively.

\subsection{Adiabatic SWAP pulses}

\begin{figure}[t]
\includegraphics[scale=0.7]{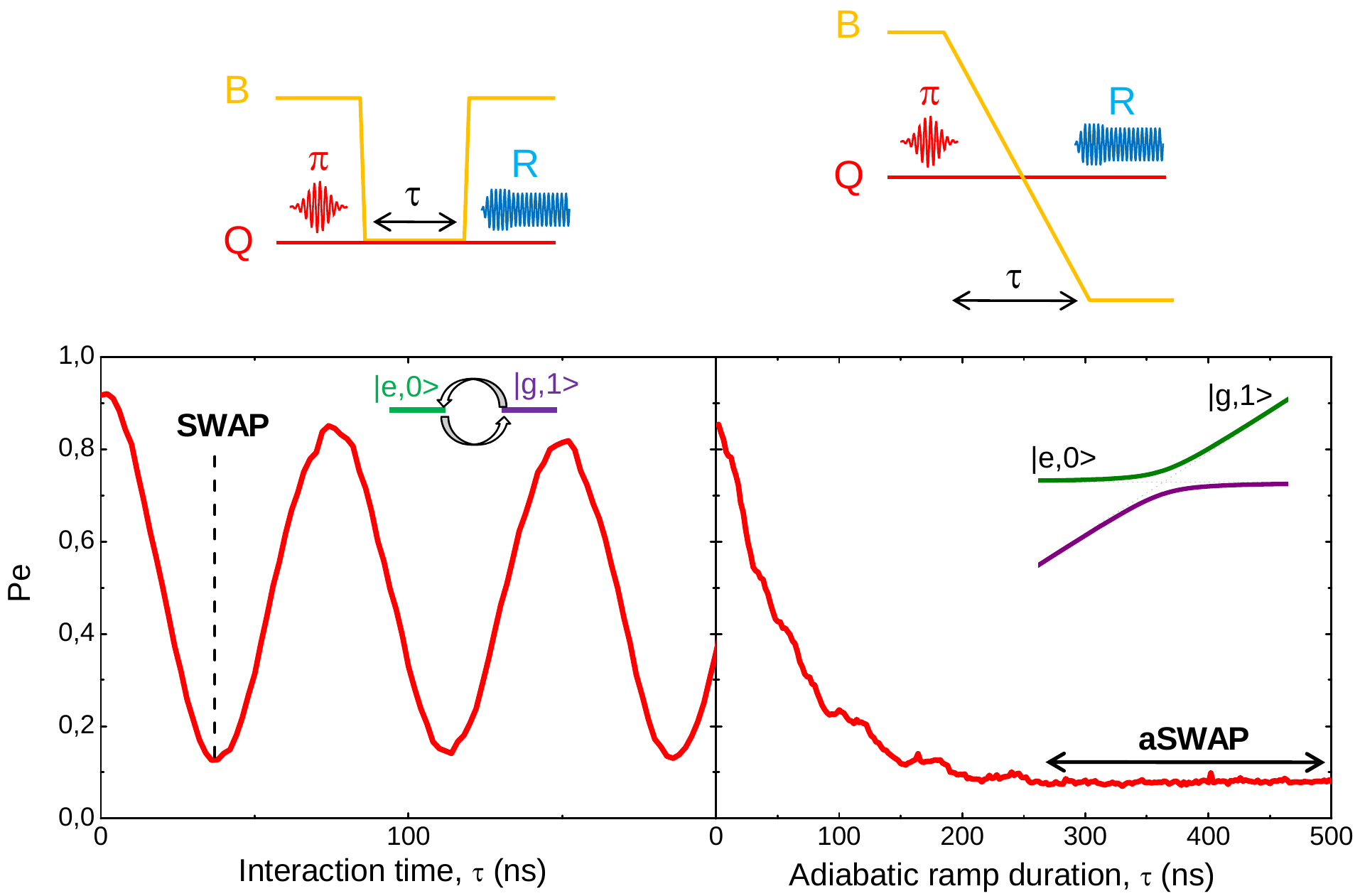}

\caption{Comparison between resonant and adiabatic SWAP pulses. Left panel:
principle of a resonant SWAP. After excitation of the qubit in $\left|e\right\rangle $,
$B$ is put suddenly into resonance with $Q$ for an interaction time
$\tau$ during which $\left|e,0\right\rangle $ and $\left|g,1\right\rangle $
exchange periodically energy. After a time $\tau=37$~ns, the qubit
excitation is transferred to $B$. Right panel: principle of an adiabatic
SWAP (aSWAP). The qubit is excited in $\left|e\right\rangle $, after
what $\omega_{B}$ is ramped through $\omega_{Q}$ in a time $\tau$,
and the state of $Q$ is finally read-out. For long enough ramp durations
(for this sequence $\tau\gtrsim300$~ns), the qubit excited state
population is fully transferred into the bus. \label{figS5:adiabatic_passage}}
\end{figure}

SWAP operations between the qubit and the resonator quantum bus can
be performed by tuning suddenly $\omega_{B}$ in resonance with $\omega_{Q}$
for a duration $\pi/2g_{Q}$. We show in Fig. \ref{figS5:adiabatic_passage}
the resulting vacuum Rabi oscillations. In the experiment however,
we found out that such a resonant SWAP operation was not stable enough
to allow subsequent data acquisition longer than $\sim15$ minutes.
The problem is caused by flux noise in the SQUID loop which causes
$\omega_{B}(\Phi)$ to change over time so that the amplitude of flux
pulse needed to perform the vacuum Rabi oscillations in resonance
also changes in time. We note that we found a much larger flux noise
for $B_{NV}=1.4$~mT than $B_{NV}=0$~mT, probably due to vortices
being trapped in the superconducting thin films around the SQUID. 

To circumvent this problem, we resort instead to adiabatic SWAP operations
in which $\omega_{B}$ is adiabatically ramped through resonance with
$\omega_{Q}$ so that state $\left|e,0\right\rangle $ is adiabatically
converted into $\left|g,1\right\rangle $ for a sufficiently slow
flux pulse, yielding the same operation as the resonant SWAP (see
Fig. \ref{figS5:adiabatic_passage}). Finding good parameters for
the pulse requires some optimization since a too fast pulse will not
be adiabatic while a too slow pulse will strongly reduce the signal
because of energy relaxation either in the qubit or in the resonator
bus. The final parameters that we used are: $\omega_{B}$ starts at
$2.52$~GHz, is first ramped up to $2.589$~GHz in $60$~ns, then
to $2.643$~GHz in $350$~ns, then to $2.687$~GHz in $40$~ns
(with a qubit frequency $\omega_{Q}/2\pi=2.607$~GHz).

\subsection{Theory}

We now explain in more detail how the theory curves in figures 2 and
4 of the article are calculated. The calculation assumes that the
qubit state is perfectly transferred to the resonator bus $B$, so
that the measured $P_{e}$ perfectly maps the final resonator population
in the $\left|1_{B}\right\rangle $ state. Each result of the calculation
is rescaled in amplitude and offset to match the experimental data
(this accounts for the additional losses caused by relaxation of the
qubit or resonator during the pulse sequence, and in particular during
the two $aSWAP$s). Apart from that, all the calculations are performed
using the following theory, and one unique parameters set for the
whole paper.

\subsubsection{Rabi oscillations (figure 2)}

What needs to be calculated is the final probability to find a photon
in the bus resonator after the resonator-spins interaction, assuming
the resonator is in $\left|1_{B}\right\rangle $ at time $\tau=0$.
The calculations are performed in the Holstein-Primakoff approximation,
in which the spins and the resonator are described by harmonic oscillators,
as explained in the Methods section. The system Hamiltonian is $H/\hbar=\omega_{B}(\Phi)a^{\dagger}a+\sum\omega_{\mathrm{j}}b_{\mathrm{j}}^{\dagger}b_{\mathrm{j}}+\sum ig_{\mathrm{j}}(b_{\mathrm{j}}^{\dagger}a-b_{\mathrm{j}}a^{\dagger})$,
$g_{\mathrm{j}}$ being the coupling constant of spin $j$ with the
resonator. We need to calculate $p(t)=\left|\alpha(t)\right|^{2}$
with $\alpha(t)=\left\langle 0\right|a(t)a^{\dagger}(0)\left|0\right\rangle $,
which represents the probability that a photon created at $t=0$ is
still present at time $t$. As shown in \cite{Diniz2011-1} this quantity
can be calculated by considering an effective non-Hermitian Hamiltonian 

\begin{equation} H_{eff}/\hbar = \left( \begin{array}{cccc} \tilde\omega_0 & i g_1 & i g_2 & \ldots \\ -i g_1 & \tilde\omega_1 & & \\ -i g_2 & & \tilde\omega_2 &\\ \vdots & & & \ddots \\ \end{array} \right)\, . \end{equation}

with complex angular frequencies $\tilde{\omega}_{B}=\omega_{B}-i\kappa/2$
and $\tilde{\omega}_{k}=\omega_{k}-i\gamma_{0}/2$ ; here, $\gamma_{0}$
is the spontaneous emission rate of each spin, and $\kappa=\omega_{B}/Q$
is the bus resonator damping rate (where we introduced its quality
factor $Q$). Indeed, introducing the vector $X(t)$ of coordinates
$\left[\left\langle a(t)a^{\dagger}(0)\right\rangle ,...,\left\langle b_{j}(t)a^{\dagger}(0)\right\rangle ,...\right]$
it can be shown that $dX/dt=-(i/\hbar)H_{eff}X$. The formal solution
to this equation is then 

\begin{equation} \label{eq:2} X(t) = \mathcal{L}^{-1}  [ (s+i H_{eff} /\hbar)^{-1}  X(0)] \, , \end{equation}

with $X(0)=x_{G}$ and $x_{G}\equiv(1,0,...,0)$ . This implies that
$\alpha(t)=x_{G}{}^{\dagger}\cdot X(t)=\mathcal{L}^{-1}\left[t_{1}(s)\right]$
with $t_{1}(s)=x_{G}{}^{\dagger}\cdot(s+iH_{eff})^{-1}\cdot x_{G}$
and $\mathcal{L}[f(s)]=\int e^{-st}f(t)dt$, $s$ being a complex
number. Since $t_{1}(s)$ is not singular on its imaginary axis, we
only need $t_{1}$ for pure imaginary argument $s=-i\omega$ to perfom
the Laplace transform inversion. As shown in \cite{Diniz2011-1},
we have $t_{1}(-i\omega)=i/\left[\omega-\omega_{B}+i\kappa/2-W(\omega)\right]$
with $W(\omega)=g^{2}\int\rho(\omega')d\omega'/\left[\omega-\omega'+i\gamma_{0}/2\right]$.
In this last formula, $g$ is the coupling constant of the spin ensemble
to the bus resonator and $\rho(\omega)$ is the density of spins which
is taken as explained below. Computing $\alpha(t)$ is thus achieved
by evaluating $t_{1}$ for the distribution $\rho(\omega)$, and numerically
evaluting the inverse Laplace transform. At the end of the calculation,
we take the $\gamma_{0}\rightarrow0$ limit since NV centers at low
temperature have negligible energy relaxation.

\subsubsection{Single-photon Ramsey experiment (figure 4b)}

For the Ramsey-like experiment (figure 4b), each $\pi/2$ pulse is
realised by bringing the resonator and spins to resonance. For a fast
pulse, the resonant interaction maps continuously the coherent state
of the field to the superradiant mode in the spins. We calibrate the
interaction time in such a way as to transform the state $x_{G}$
into the superposition $\frac{x_{G}-x_{S}}{\sqrt{2}}$. After the
first $\pi/2$ pulse, the resonator is kept detuned from the spin
ensemble for a time $t$. The system state at this point can be evaluated
using eq. (2). We define $X_{G}(t)$ (resp. $X_{S}(t)$) as the vector
of coordinates $\left[\left\langle a(t)a^{\dagger}(0)\right\rangle ,...,\left\langle b_{j}(t)a^{\dagger}(0)\right\rangle ,...\right]$
at time $t$ with initial conditions $x_{G}$ (resp. $x_{S}$). A
second $\pi/2$ pulse is then applied before the amplitude $\alpha(t)$
of the field in the resonator is measured:

\begin{equation}     \alpha(t) = \frac{1}{\sqrt 2} x_G^\dagger \cdot U_{\pi/2} (X_G(t) - X_S(t) )  \\ 
  = \frac{1}{2} (x_G^\dagger + x_S^\dagger ) \cdot (X_G(t) - X_S(t) ) \\% 
  =\frac{1}{2} \mathcal{L}^{-1} ( t_1(s) - t_2(s) + t_3(s) - t_4(s) ) \, \end{equation} where $t_{2}(s)=x_{S}{}^{\dagger}\cdot(s+iH_{eff})^{-1}\cdot x_{S}$
and $t_{2}(s)=x_{S}{}^{\dagger}\cdot(s+iH_{eff})^{-1}\cdot x_{G}$,
$t_{1}$ and $t_{4}$ are defined above. As shown in \cite{Diniz2011-1},
$t_{2}(-i\omega)=-t_{1}(-i\omega)W(\omega)(s+i\tilde{\omega}_{0})/g^{2}$
and $t_{3}=-t_{4}$.

\subsubsection{Parameters used in the simulation}

We thus see that the only thing that is needed to numerically perform
these calculations is the density of spins $\rho(\omega)$ and the
ensemble coupling constant $g$. The spin density $\rho(\omega)$
is chosen as the sum of three Lorentzian peaks separated by $2.3$~MHz.
The only parameters that are adjusted in order to fit the data are
the coupling constants and the peak linewidth in the HF structure.
As explained in the main text, we find that $g_{I}/2\pi=2.9$~MHz
and $g_{III}/2\pi=3.8$~MHz fit best our data, with a linewidth of
$1.6$~MHz for the spins belonging to group $I$, and $2.4$~MHz
for the spins belonging to group $III$. We attribute the larger linewidth
of the spins from group $III$ to a residual misalignment of $B_{NV}$
with respect to the $[1,1,1]$ axis of the crystal which causes each
of the three $<1,1,1>$ axes non-collinear with the field to undergo
slightly different Zeeman shifts. A misalignment of $0.02$~rad would
be enough to cause a broadening such as we observe.

We also note that the splitting of $2.3$~MHz between the three peaks
of the HF structure is slightly larger than the value reported in
most articles which is $2.18$~MHz. Our data are however not precise
enough to determine precisely whether this difference actually reflects
a change in the HF interaction parameters of the NV center at low
temperature.

\end{document}